\documentclass[english,twocolumn,prb,superscriptaddress]{revtex4-2}
\usepackage{amssymb}
\usepackage{enumerate}
\usepackage{amsmath}
\usepackage{bbm}
\usepackage{color}
\usepackage{tikz}
\usepackage{appendix}
\usepackage{hyperref}
\hypersetup{
  colorlinks=true,
  citecolor=blue,
  linkcolor=blue,
  urlcolor=blue,
}

\begin{document}
\title{Dynamical detection of mean-field topological phases in an interacting Chern insulator}
\author{Wei Jia}
\affiliation{International Center for Quantum Materials and School of Physics, Peking University, Beijing 100871, China}
\affiliation{Collaborative Innovation Center of Quantum Matter, Beijing 100871, China}
\author{Long Zhang}
\affiliation{School of Physics and Institute for Quantum Science and Engineering, Huazhong University of Science and Technology, Wuhan 430074, China}
\affiliation{International Center for Quantum Materials and School of Physics, Peking University, Beijing 100871, China}
\affiliation{Collaborative Innovation Center of Quantum Matter, Beijing 100871, China}
\author{Lin Zhang}
\affiliation{ICFO-Institut de Ciencies Fotoniques, The Barcelona Institute of Science and Technology, Av. Carl Friedrich Gauss 3, 08860 Castelldefels (Barcelona), Spain}
\affiliation{International Center for Quantum Materials and School of Physics, Peking University, Beijing 100871, China}
\affiliation{Collaborative Innovation Center of Quantum Matter, Beijing 100871, China}
\author{Xiong-Jun Liu}
\thanks{Correspondence addressed to: xiongjunliu@pku.edu.cn}
\affiliation{International Center for Quantum Materials and School of Physics, Peking University, Beijing 100871, China}
\affiliation{Collaborative Innovation Center of Quantum Matter, Beijing 100871, China}
\affiliation{CAS Center for Excellence in Topological Quantum Computation, University of Chinese Academy of Sciences, Beijing 100190, China}
\affiliation{Institute for Quantum Science and Engineering and Department of Physics, Southern University of Science and Technology, Shenzhen 518055, China}


\begin{abstract}
Interactions generically have important effects on the topological quantum phases. For a quantum anomalous Hall (QAH) insulator, the presence of interactions can qualitatively change the topological phase diagram which, however, is typically hard to measure in the experiment. Here we propose a novel scheme based on quench dynamics to detect the mean-field topological phase diagram of an interacting Chern insulator described by QAH-Hubbard model, with nontrivial dynamical quantum physics being uncovered. We focus on the dynamical properties of the system at a weak to intermediate Hubbard interaction which mainly induces a ferromagnetic order under the mean-field level. {\color{black}Remarkably, three characteristic times $t_s$, $t_c$, and $t^*$ are found in the quench dynamics. The first two capture the emergence of dynamical self-consistent particle density and dynamical topological phase transition respectively, while the last one gives a linear scaling time on the topological phase boundaries. A more interesting result is that $t_s>t^*>t_c$ ($t^*<t_s<t_c$) occurs in repulsive (attractive) interaction and the Chern number is determined by any two characteristic time scales when the system is quenched from an initial nearly fully polarized state to the topologically nontrivial regimes, showing a dynamical way to determine equilibrium mean-field topological phase diagram via the time scales. Experimentally, 
the measurement of $t_s$ is challenging while $t_c$ and $t^*$ can be directly readout by measuring the spin polarizations of four Dirac points and the time-dependent particle density, respectively. Our work reveals the novel interacting effects on the topological phases and shall promote the experimental observation.}
\end{abstract}

\maketitle

\section{Introduction}
Topological quantum phase is currently a mainstream of research in condensed matter physics~\cite{klitzing1980new,tsui1982two,hasan2010colloquium,wen2017colloquium,kosterlitz2017nobel}. At equilibrium, the topological phases can be characterized by nonlocal topological invariants~\cite{wen1990topological,qi2011topological} defined in ground states. This classifies the gapped band structures into distinct topological states, with great success having been achieved in study of topological insulators~\cite{fu2007topological,fu2007topologicalPRB,fu2011topological,chang2013experimental}, topological semimetals~\cite{burkov2011weyl,young2012dirac,xu2015discovery,lv2015experimental}, and topological superconductors~\cite{qi2009time,bernevig2013topological,ando2015topological,sato2017topological,zhang2018observation}. Nevertheless, this noninteracting topological phase can be greatly affected after considering many-body interaction~\cite{wang2010topological,pesin2010mott,lee2011effects,castro2011topological,araujo2013change,yao2013interaction,messer2015exploring,spanton2018observation,rachel2018interacting,viyuela2018chiral,andrews2020fractional,mook2021interaction}. For instance, the repulsive Hubbard interaction can drive a trivial insulator into a topological Mott insulator~\cite{raghu2008topological,rachel2010topological,vanhala2016topological}, while the attractive Hubbard interaction may drive a trivial phase of two-dimensional (2D) quantum anomalous Hall (QAH) system into a topological superconductor/superfluid~\cite{qi2006topological,qi2010chiral,liu2014realization,poon2018semimetal}. Hence how to accurately identify the topological phases driven by the interactions is still a fundamental issue and usually hard in experiments.

In recent years, the rapid development of quantum simulations~\cite{hofstetter2018quantum,fauseweh2021digital,monroe2021programmable}
provides new realistic platforms to explore exotic interacting physics, such as ultracold atoms in optical lattices~\cite{bloch2008many,langen2015ultracold,gross2017quantum,schreiber2015observation,smith2016many,bordia2016coupling,choi2016exploring} and superconducting qubits~\cite{ramasesh2017direct,jiang2018quantum,yan2019strongly}. A number of topological models have been realized in experiments, such as the 1D Su-Schriffer-Heeger model~\cite{su1980soliton,atala2013direct}, 1D AIII class topological insulator~\cite{liu2013manipulating,song2018observation}, 1D bosonic symmetry-protected phase~\cite{haldane1983continuum,de2019observation}, 2D Haldane model~\cite{jotzu2014experimental}, the spin-orbit coupled QAH model~\cite{wu2016realization,sun2018highly,liang2021realization}, and the 3D Weyl semimetal band~\cite{he2016realization,wang2018dirac,lu2020ideal,wang2021realization,li2021weyl}. Accordingly, the various detection schemes for the exotic topological physics are also developed, ranging from the measurements of equilibrium topological physics~\cite{liu2013detecting,hafezi2014measuring,wu2014topological,price2016measurement} to non-equilibrium quantum dynamics~\cite{vajna2015topological,hu2016dynamical,budich2016dynamical,wilson2016remnant,heyl2018dynamical,heyl2019dynamical,hu2020dynamical,hu2020topological,kuo2021decoherent,cai2022synthetic}. In particular, the dynamical characterization~\cite{zhang2018dynamical,zhang2019characterizing,ye2020emergent,jia2021dynamically,li2021direct,zhang2021universal,fang2022generic} shows the correspondence between broad classes of equilibrium topological phases and the emergent dynamical topology in far-from-equilibrium quantum dynamics induced by quenching such topological systems, which brings about the systematic and high-precision schemes to detect the topological phases based on quantum dynamics and has advanced broad studies in experiment~\cite{sun2018uncover,tarnowski2019measuring,yi2019observing,wang2019experimental,xin2020quantum,ji2020quantum,niu2020simulation,chen2021digital,yu2021topological,zhang2022topological}. Nevertheless, these current studies have been mainly focused on the noninteracting topological systems, while the particle-particle interactions are expected to have crucial effects on the topological phases, whose detection is typically hard to achieve. For example, when quenching an interacting topological system~\cite{manmana2007strongly,polkovnikov2011colloquium,kiendl2017many,zhang2021nonequilibrium}, both the interactions and many-body states of the system evolve simultaneously after quenching, leading to complex nonlinear quantum dynamics~\cite{moeckel2008interaction,han2012evolution,foster2013quantum,dong2015dynamical} and exotic nonequilibrium phenomena~\cite{gornyi2005interacting,eisert2015quantum,yao2017discrete,peotta2021determination}. With above considerations, there are two nontrivial issues for the quantum dynamics in the interacting Chern insulator that have not been studied: (i) how the dynamical properties change when the Hubbard interaction is added into the Chern insulator and performing quench dynamics? (ii) is there any universal feature in the dynamical evolution to characterize the equilibrium mean-field topological phases?

In this paper, we address these issues and propose a novel scheme based on quench dynamics to detect the mean-field topological phase diagram of an interacting Chern insulator described by QAH-Hubbard model, with nontrivial dynamical quantum physics being uncovered. Specifically, we consider a 2D QAH system in presence of a weak to intermediate Hubbard interaction which mainly induces a ferromagnetic order under the mean-field level. By quenching the system from an initial nearly fully polarized trivial state to a parameter regime in which the equilibrium phase is topologically nontrivial, we uncover two dynamical phenomena rendering the dynamical signals of the equilibrium mean-field phase. {\color{black}First, there are three characteristic times $t_s$, $t_c$ , and $t^*$, capturing the dynamical self-consistent particle density, dynamical topological phase transition, and the linear scaling time on the topological phase boundaries, respectively. Second, $t_s>t^*>t_c$ ($t^*<t_s<t_c$) occurs in the repulsive (attractive) interaction and the Chern number is determined by any two characteristic time scales. Based on these two fundamental properties, we can easily determine the equilibrium mean-field topological phase diagram by comparing any two time scales. Experimentally, the measurement of $t_s$ is challenging while $t_c$ and $t^*$ can be directly readout by measuring the spin polarizations of four Dirac points and the time-dependent particle density respectively. This result provides a dynamical detection scheme with high feasibility and simplicity for measuring the mean-field topological phase diagram in interacting systems, which may be applied to the recent quantum simulation experiments.} 

The remaining part of this paper is organized as follows.
In Sec.~\ref{sec:QAH-Hubbard model}, we introduce the QAH-Hubbard model.
In Sec.~\ref{sec:Quench dynamics}, we study the quench dynamics of the system.
In Sec.~\ref{sec:Nontrivial dynamical properties}, we reveal the nontrivial dynamical properties in quench dynamics. 
In Sec.~\ref{sec:Mean-field phase diagram determined by time scales}, we determine the mean-field topological phase diagram via the time scales.
In Sec.~\ref{sec:Experimental detection}, we propose the experimental detection scheme for the equilibrium mean-field topological phases. Finally, we summarize the main results and provide the brief discussion in Sec.~\ref{sec:Conclusion}.

\section{QAH-Hubbard model}\label{sec:QAH-Hubbard model}
Our starting point is a minimal 2D QAH model~\cite{qi2006topological,liu2014realization}, which has been recently realized in cold atoms~\cite{wu2016realization,sun2018uncover,yi2019observing,liang2021realization}, together with an attractive or repulsive on-site Hubbard interaction of strength $U$. The system is now described by the QAH-Hubbard Hamiltonian
\begin{equation}\label{eq:1}
\begin{aligned}
H & =\sum_{\mathbf{k}}C^{\dag}_{\mathbf{k}}\mathcal{H}^{(0)}_{\mathbf{k}}C_{\mathbf{k}}+U\sum_{\mathbf{j}}n_{\mathbf{j}\uparrow}n_{\mathbf{j}\downarrow},\\
\mathcal{H}^{(0)}_{\mathbf{k}} & = \mathbf{h}_{\mathbf{k}}\cdot\boldsymbol{\sigma} =\left[m_z-2t_0(\cos k_x+\cos k_y)\right]\sigma_z\\
&\quad+2t_{\text{so}}\sin k_y\sigma_x+2t_{\text{so}}\sin k_x\sigma_y,
\end{aligned}
\end{equation}
where $C_{\mathbf{k}}=(c_{\mathbf{k}\uparrow},c_{\mathbf{k}\downarrow})^{T}$ is the spinor operator of momentum $\mathbf{k}$, $n_{\mathbf{j}s}=c^{\dagger}_{\mathbf{j}s}c_{\mathbf{j}s}$ with $s=\uparrow$ or $\downarrow$ is the particle number operator at site $\mathbf{j}$, $\sigma_{x,y,z}$ are the Pauli matrices, and $m_z$ is the Zeeman coupling. Here $t_0$ and $t_{\text{so}}$ are the spin-conserved and spin-flipped hopping coefficients, respectively. In the noninteracting case, the Bloch Hamiltonian $\mathcal{H}^{(0)}_{\mathbf{k}}$ produces two energy bands $\pm e_{\mathbf{k}}=\pm\sqrt{h_{x,\mathbf{k}}^2+h_{y,\mathbf{k}}^2+h_{z,\mathbf{k}}^2}$, for which the gap can be closed at Dirac points $\mathbf{D}_i\in\{\mathbf{X}_1,\mathbf{X}_2,\boldsymbol{\Gamma},\mathbf{M}\}$ with $\mathbf{X}_1=(0,\pi)$, $\mathbf{X}_2=(\pi,0)$, $\boldsymbol{\Gamma}=(0,0)$, and $\mathbf{M}=(\pi,\pi)$ for certain Zeeman coupling. When the system is fully gapped, the corresponding band topology can be characterized by the first Chern number $\text{Ch}_1$, determining the QAH topological region $0<|m_z|<4t_0$ with $\text{Ch}_1=\text{sgn}(m_z)$ and the trivial region $|m_z|>4t_0$. This noninteracting topological property can also be captured by the intuitive physical quantities, such as the numbers of edge states~\cite{qi2006topological}, the spin textures on band inversion surfaces~\cite{zhang2018dynamical,yi2019observing}, and the spin polarizations at four Dirac points~\cite{liu2013detecting,wu2016realization}.

The presence of nonzero interactions can greatly affect the physics of Hamiltonian \eqref{eq:1} and may induce various interesting quantum phases at suitable interaction strength, such as the superfluid phase in attractive interaction~\cite{jia2019topological,powell2022superfluid} and the antiferromagnetic phase in repulsive interaction~\cite{esslinger2010fermi,tarruell2018quantum}. Especially for a general strong interaction, the system may host rich magnetic phases~\cite{ziegler2020correlated,ziegler2022large,tirrito2022large}. Here we focus on the 2D system at half filling within a weak to intermediate interacting regime, where the interaction mainly induces ferromagnetic order~\cite{guo2011topological}. For this 2D case, the Hubbard interaction can be still treated by employing the mean field theory, in which the fluctuations around the average value of the order parameter are small and can be neglected~\cite{qi2010chiral,liu2014realization,poon2018semimetal,han2012evolution,foster2013quantum,dong2015dynamical}. Accordingly, the Hubbard interaction term is rewritten as
\begin{equation}
\sum_{\mathbf{j}}n_{\mathbf{j}\uparrow}n_{\mathbf{j}\downarrow}=n_{\uparrow}\sum_{\mathbf{k}}c_{\mathbf{k} \downarrow}^{\dagger} c_{\mathbf{k} \downarrow}+n_{\downarrow}\sum_{\mathbf{k}}c_{\mathbf{k} \uparrow}^{\dagger} c_{\mathbf{k} \uparrow}-Nn_{\uparrow}n_{\downarrow}
\end{equation}
with $n_{s}=({1}/{N})\sum_{\mathbf{k}} \langle c_{\mathbf{k}s}^{\dagger} c_{\mathbf{k}s}\rangle$. Here $N$ is the total number of sites. It is clear that the nonzero interaction corrects the Zeeman coupling to an effective form
\begin{equation}\label{m_eff}
m^{\text{eff}}_z=m_z-U\frac{(n_{\uparrow}-n_{\downarrow})}{2}=m_z-Un_d,
\end{equation}
where $n_d\equiv (n_\uparrow-n_\downarrow)/2$ is the difference of density for spin-up and spin-down particles. The effective Zeeman coupling shifts the topological region to $0<|m^{\text{eff}}_z|<4t_0$ with $\text{Ch}_1=\text{sgn}(m^{\text{eff}}_z)$ in the interacting regime.

\begin{figure}[t]
\includegraphics[width=\columnwidth]{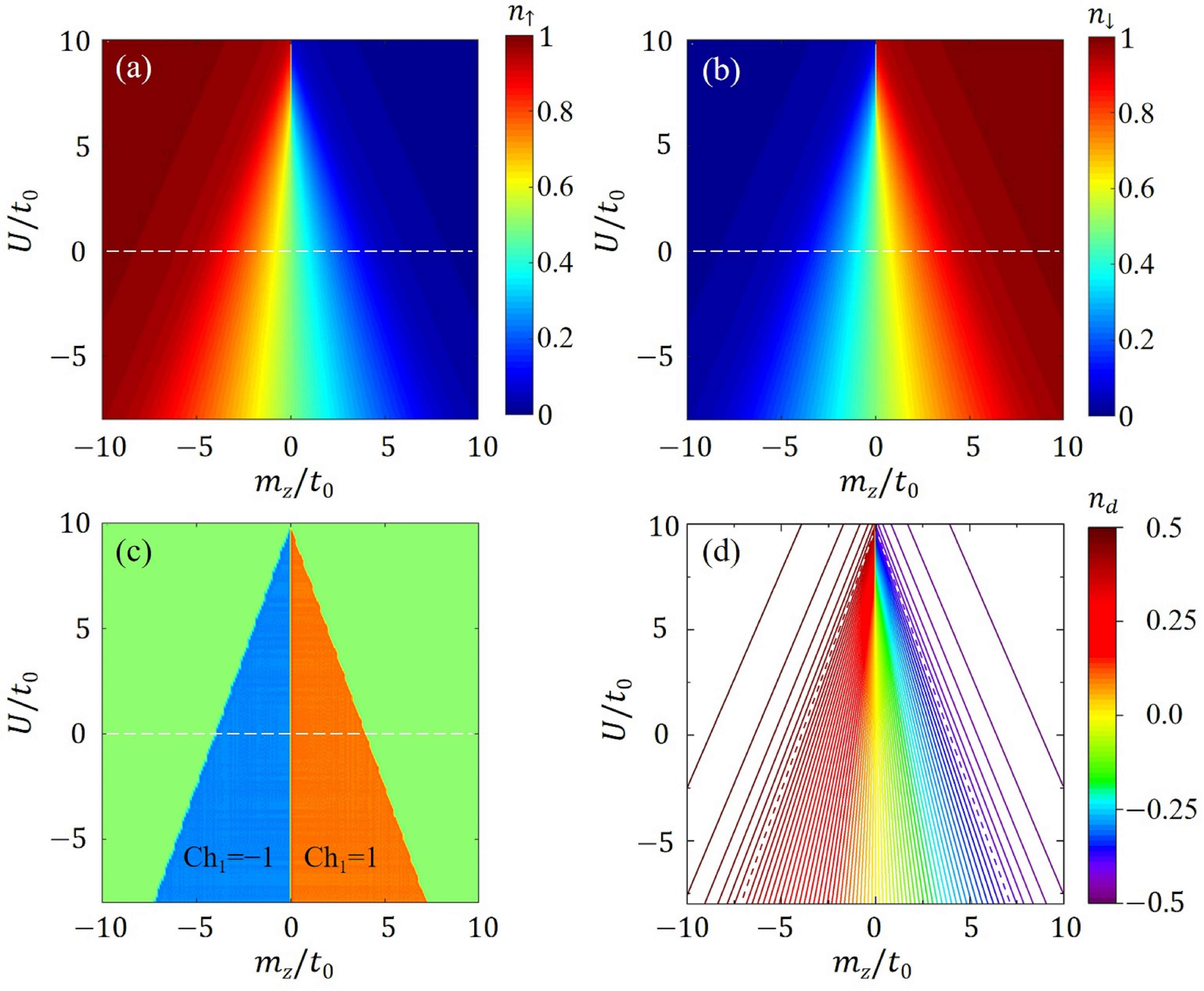}
\caption{(a)-(b) Self-consistent results of $n_{\uparrow}$ and $n_{\downarrow}$.
(c)~Mean-field phase diagram for the nonzero $U$ and $m_z$. (d)~Contour lines of $n_d$, where the topological phase boundaries (two dashed lines) have $n^{*}_d=\pm 0.4068$. Here we set $t_{\text{so}}=t_0$ and $N=60\times 60$.}
\label{Fig1}
\end{figure}

By self-consistently calculating the particle density $n_{\uparrow (\downarrow)}$ [see Figs.~{\ref{Fig1}}(a) and {\ref{Fig1}}(b)], the mean-field phase diagram of the QAH-Hubbard model \eqref{eq:1} is shown in Fig.~{\ref{Fig1}}(c). We find an important feature that the topological phase boundaries depend on the strength of nonzero interaction and the Zeeman coupling linearly~[see Appendix \ref{appendix-1}]. Indeed, this linear behavior of topological boundaries is essentially a natural consequence of the linear form of the contour lines of $n_d$ in terms of the interaction strength and Zeeman coupling, as we show in Fig.~{\ref{Fig1}}(d). On these lines, $m^{\mathrm{eff}}_{z}$ is also unchanged, since $n_{d}$ is fully determined by the effective Zeeman coupling. Specifically, we have $n^{*}_d=\pm 0.4068$ on the topological phase boundaries $|m^{\mathrm{eff}}_{z}|=4t_{0}$ for $t_{\mathrm{so}}=t_{0}$. This linear scaling form of $n_{d}$ greatly affects the dynamical properties of the system and leads to novel phenomena in the quench dynamics, as we show in Sec.~\ref{sec:Nontrivial dynamical properties}. 

\section{Quench dynamics}\label{sec:Quench dynamics}

Quantum quench dynamics has been widely used in cold atoms~\cite{sun2018highly,sun2018uncover,song2019observation,wang2021realization}. We shall show that the above equilibrium mean-field phase diagram can be dynamically characterized and detected by employing the quantum quench scheme. Unlike the interaction quench in Refs.~\cite{moeckel2008interaction,foster2013quantum}, here we choose the Zeeman coupling as quench parameter, which has the following advantages: (i) The effective Zeeman coupling directly determines the topology of mean-field ground states at equilibrium; (ii) The Zeeman field in spin-orbit coupled quantum gases is controlled by the laser intensity and/or detuning and can be changed in a very short time scale, fulfilling the criterion for a sudden quench; (iii) The realistic experiments~\cite{lin2011spin,wang2012spin,williams2013raman} have demonstrated that both the magnitude and the sign of Zeeman field can be tuned, which is more convenient to operate. 

The quench protocol is as follows. First, we initialize the system into a nearly fully polarized state for time $t<0$ by taking a very large constant magnetization $m^{(c)}_z$ along the $\sigma_{z}$ axis but a very small $m^{(c)}_{x(y)}$ along $\sigma_{x(y)}$ axis. In this case the effect of interaction $U$ can be ignored, and the spin of initial state is almost along the $z$ axis with only a very small component in the $x$-$y$ plane. At $t=0$, we suddenly change the Zeeman coupling $m^{(c)}_z$ to the post-quenched value $m_z$ and remove $m^{(c)}_{x(y)}$. And then, the nearly fully polarized state begins to evolve under the equation of motion $i \dot{\Psi}_{\mathbf{k}}(t)=\mathcal{H}_{\mathbf{k}}(t)\Psi_{\mathbf{k}}(t)$ with the post-quenched Hamiltonian
\begin{equation}\label{eq:2}
\mathcal{H}_{\mathbf{k}}(t)={\left[ \begin{array}{cc}
h_{z,\mathbf{k}}+Un_\downarrow(t) & h_{x,\mathbf{k}}-ih_{y,\mathbf{k}} \\
h_{x,\mathbf{k}}+ih_{y,\mathbf{k}} & -h_{z,\mathbf{k}}+Un_{\uparrow}(t)
\end{array}
\right]},
\end{equation}
where $\Psi_{\mathbf{k}}(t)=[\chi_{\mathbf{k}}(t),\eta_{\mathbf{k}}(t)]^T$ is instantaneous many-body state. We emphasize $m^{(c)}_z$ should have the same sign as the post-quenched $m_z$ in order to facilitate the capture of nontrivial properties in quantum dynamics. Also, all physics in the dynamics are robust against the tiny changes $|m^{(c)}_{x(y)}|\in[0,m_z]$ of the initial state. 

In the progress of quench dynamics, we observe that the many-body state $\Psi_{\mathbf{k}}(t)$ and the post-quenched Hamiltonian $\mathcal{H}_{\mathbf{k}}(t)$ are both time-evolved, where the instantaneous particle density is time-dependent and determined as 
\begin{equation}\label{eq:3}
n_{\uparrow}(t)=\frac{1}{N} \sum_{\mathbf{k}}|\chi_\mathbf{k}(t)|^2,~
n_{\downarrow}(t)=\frac{1}{N} \sum_{\mathbf{k}}|\eta_\mathbf{k}(t)|^2.
\end{equation}
This dynamic behavior is completely different from the noninteracting quantum quenches, where the post-quenched Hamiltonian keeps unchanged~\cite{zhang2018dynamical,zhang2019characterizing,ye2020emergent,jia2021dynamically,li2021direct,zhang2021universal}. Even if the post-quenched Hamiltonian may become steady after the long time evolution, it is still different from the equilibrium mean-field Hamiltonian with the post-quenched $m_{z}$ and $U$, i.e.,
\begin{equation}\label{ap-1}
\mathcal{H}^{\text{MF}}_\mathbf{k}={\left[ \begin{array}{cc}
h_{z,\mathbf{k}}-Un_d+\frac{U}{2} & h_{x,\mathbf{k}}-ih_{y,\mathbf{k}} \\
h_{x,\mathbf{k}}+ih_{y,\mathbf{k}} & -h_{z,\mathbf{k}}+Un_d+\frac{U}{2}
\end{array}
\right]}.
\end{equation}
Here $n_{d}$ is the equilibrium value as we show in the Section \ref{sec:QAH-Hubbard model}. Clearly, the novel dynamical evolution shows that the detection schemes for the noninteracting topological phases are inapplicable in this interacting case. 

\begin{figure}[t]
\includegraphics[width=\columnwidth]{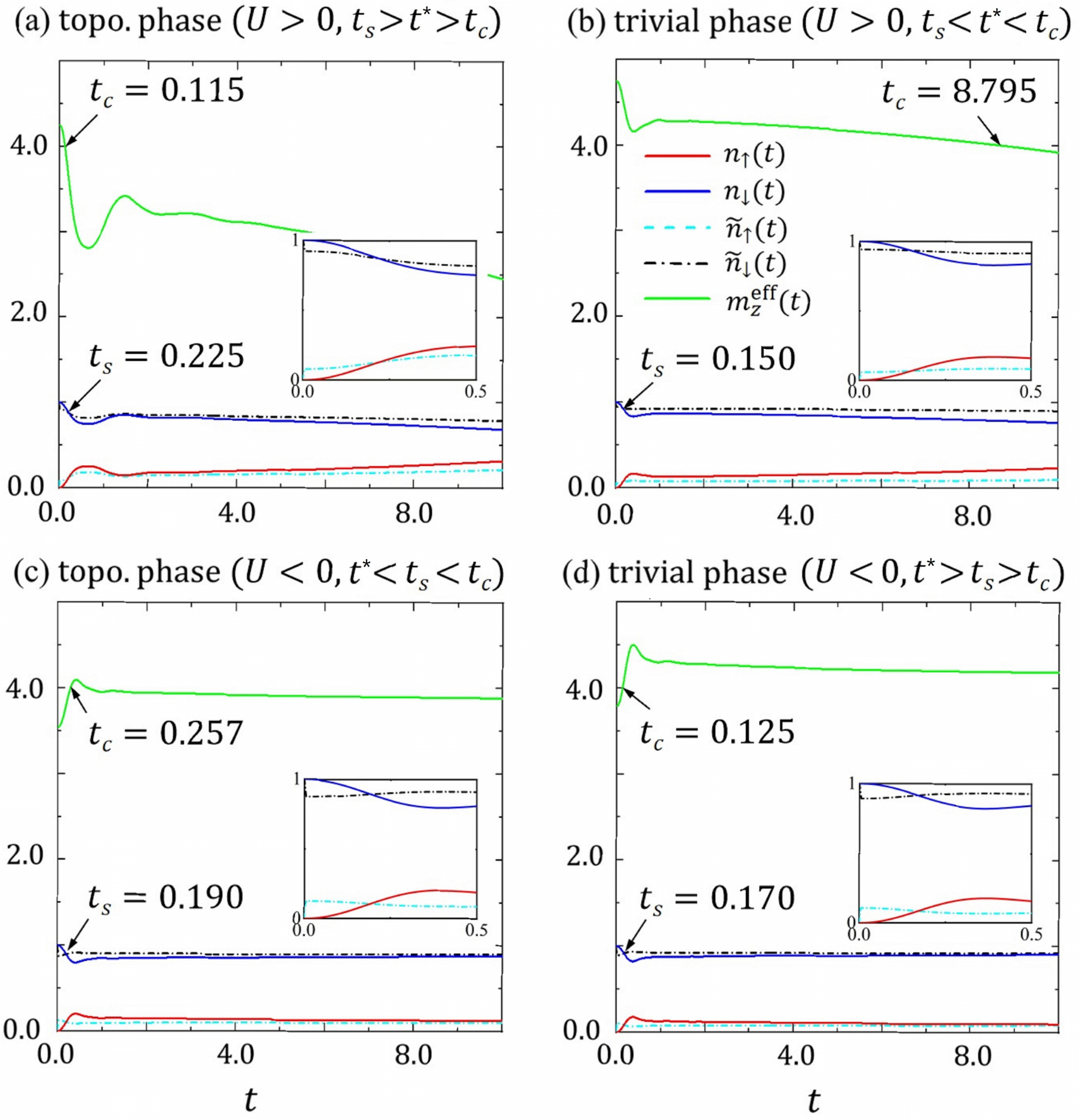}
\caption{Numerical results for $n_{\uparrow}(t)$ (red solid lines), $n_{\downarrow}(t)$ (blue solid lines), $m^{\text{eff}}_z(t)$ (green solid lines), $\tilde{n}_{\uparrow}(t)$ (light-blue dash-dotted lines), and $\tilde{n}_{\downarrow}(t)$ (black dash-dotted lines) for different interaction and equilibrium mean-field topological phases. Two characteristic times $t_s$ and $t_c$ emerge in the dynamical time evolution. Here the post-quenched parameters are $\left(m_z,U\right)=(1.4,5.7)$, $(3.0,3.5)$, $(4.9,-2.7)$, and $(5.8,-4.0)$ for (a)-(d), respectively. The other characteristic time is $t^*=0.180$. We set $t_{\text{so}}=t_0=1$, $m^{(c)}_{x}=m_z$, $m^{(c)}_y=0$, and $m^{(c)}_z=100$.
}
\label{Fig2}
\end{figure}

Next we employ the nontrivial characteristic time emerged in the above dynamical evolution to identify the mean-field topological phases. By defining the quantity $n_{d}(t)\equiv[n_{\uparrow}(t)-n_{\downarrow}(t)]/2$, the time-dependent effective Zeeman coupling is given by
\begin{equation}
m^{\text{eff}}_z(t)=m_z-Un_d(t).
\end{equation}
Their dynamical properties are shown in Fig.~\ref{Fig2}, where $n_{d}(t)$ in a short-time evolution increases (decays) from an initial $n_{d}(t=0)\approx -0.5$ ($0.5$) when the post-quenched Zeeman coupling is positive (negative). It should be noted that $m^{\text{eff}}_z(t)$ now determines the topological number $W(t)$ of post-quenched Hamiltonian $\mathcal{H}_{\mathbf{k}}(t)$ through the relation $W(t)=\text{sgn}[m^{\text{eff}}_z(t)]$ for $0<|m^{\text{eff}}_z(t)|<4t_0$ and $W(t)=0$ otherwise. Considering that the topology of equilibrium mean-field Hamiltonian \eqref{ap-1} is actually determined by its self-consistent particle density $n_{\uparrow(\downarrow)}$, a idea of capturing the equilibrium mean-field topological phases is to find a characteristic time $t_{s}$ in this dynamics such that 
\begin{equation}
n_{\uparrow(\downarrow)}(t_{s})=n_{\uparrow(\downarrow)},
\end{equation}
which gives a dynamical self-consistent particle density $n_{\uparrow(\downarrow)}(t_{s})$ to characterize the equilibrium self-consistent particle density. Now this post-quenched Hamiltonian at $t_s$ is equivalent to the equilibrium mean-field Hamiltonian, and both of them host the same ground states. Accordingly, $m^{\mathrm{eff}}_{z}(t_{s})$ determines the topology of the equilibrium mean-field Hamiltonian. 
On the other hand, we introduce a characteristic time $t_{c}$ such that
\begin{equation}
|m^{\text{eff}}_z(t_c)|=4t_0,
\end{equation}
characterizing the critical time of the dynamical topological phase transition of post-quenched Hamiltonian. {\color{black}Meanwhile, a nontrivial relation $m^{\mathrm{eff}}_{z}(t_{s})=m^{\mathrm{eff}}_{z}(t_{c})$ happens on the topological phase boundaries, giving $t_s=t_c$. We remark this characteristic time on these topological boundaries as $t^*$ such that
\begin{equation}
n_d(t^*)=n^*_d,
\end{equation}
which holds a unique value for the system with a fixed spin-orbit coupled strength due to the linear scaling of $n^*_d$ [see Fig.~\ref{Fig1}(d)].
With these three characteristic time scales $t_{s}$, $t_{c}$, and $t^*$, the properties of equilibrium mean-field Hamiltonian can be completely characterized, where the ground states, the topological phase transition, and the topological phase boundaries are characterized by $\mathcal{H}_{\mathbf{k}}(t_s)$, $\mathcal{H}_{\mathbf{k}}(t_c)$, and $\mathcal{H}_{\mathbf{k}}(t^*)$ respectively.} In Fig.~\ref{Fig2}, we show that both $t_{s}$ and $t_{c}$ emerge in the short-time evolution. Together with $t^*$ in Fig.~\ref{Fig5}(b), we clearly observe that the relative time scales behave differently for the different equilibrium mean-field phases. This result provides a basic idea to establish the topological characterization and to detect the mean-field phase diagram. Next we shall theoretically calculate the three characteristic times in the following Section \ref{sec:Nontrivial dynamical properties}, and then show how to accurately identify the equilibrium mean-field topological phases via the three time scales in the Section \ref{sec:Mean-field phase diagram determined by time scales}.

\section{Nontrivial dynamical properties}\label{sec:Nontrivial dynamical properties}

In this section, we determine the characteristic times $t_{s}$, $t_{c}$, and $t^*$ entirely from quantum dynamics, and provide the analytical results under certain conditions. Also, the nontrivial dynamical properties are uncovered, including the emergence of dynamical self-consistent particle density and dynamical topological phase transition and then the scaling properties of the three characteristic times.

\subsection{Dynamical self-consistent particle density and characteristic time $t_s$}

We first figure out how to theoretically determine the characteristic time $t_{s}$ in the quantum dynamics. A notable fact is that the equilibrium particle density $n_{\uparrow(\downarrow)}$ is self-consistently obtained in the equilibrium mean-field Hamiltonian, and the time-evolved particle density $n_{\uparrow(\downarrow)}(t)$ can be considered as a specific path for updating the mean-field parameters. Hence we introduce another set of time-dependent particle density
\begin{equation}\label{eq:4}
\tilde{n}_{\uparrow}(t)=\frac{1}{N} \sum_{\mathbf{k}}|\tilde{\chi}_\mathbf{k}(t)|^2,\quad
\tilde{n}_{\downarrow}(t)=\frac{1}{N} \sum_{\mathbf{k}}|\tilde{\eta}_\mathbf{k}(t)|^2
\end{equation}
for the eigenvector $\tilde{\Psi}_{\mathbf{k}}(t) =[\tilde{\chi}_{\mathbf{k}}(t),\tilde{\eta}_{\mathbf{k}}(t)]^T$ of the post-quenched Hamiltonian $\mathcal{H}_{\mathbf{k}}(t)$ with negative energy, which can be obtained once we know the instantaneous particle density $n_{\uparrow(\downarrow)}(t)$. Note that $n_{\uparrow(\downarrow)}(t)$ and $\tilde{n}_{\uparrow(\downarrow)}(t)$ are slightly different, where the former is the particle density corresponding to the instantaneous states but the latter is that one corresponding to the eigenstates of $\mathcal{H}_{\mathbf{k}}(t)$. We see that $\mathcal{H}_{\mathbf{k}}(t)$ is self-consistently when the two sets of time-dependent particle density coincide, giving the characteristic time
\begin{equation}\label{eq:5}
t_s\equiv \text{min}\{t|n_{\uparrow(\downarrow)}(t)=\tilde{n}_{\uparrow(\downarrow)}(t)\},
\end{equation}
at which there is $n_{\uparrow(\downarrow)}=n_{\uparrow(\downarrow)}(t_s)=\tilde{n}_{\uparrow(\downarrow)}(t_s)$, as we show in Fig.~\ref{Fig2}. This result determines the dynamical self-consistent particle density $n_{\uparrow(\downarrow)}(t_{s})$ and captures the equilibrium self-consistent particle density. {\color{black}Although the measurement of $t_s$ is almost impossible in the recent experiments, it clearly characterizes the novel dynamic properties derived from quenching.} 

\begin{figure}[t]
\includegraphics[width=\columnwidth]{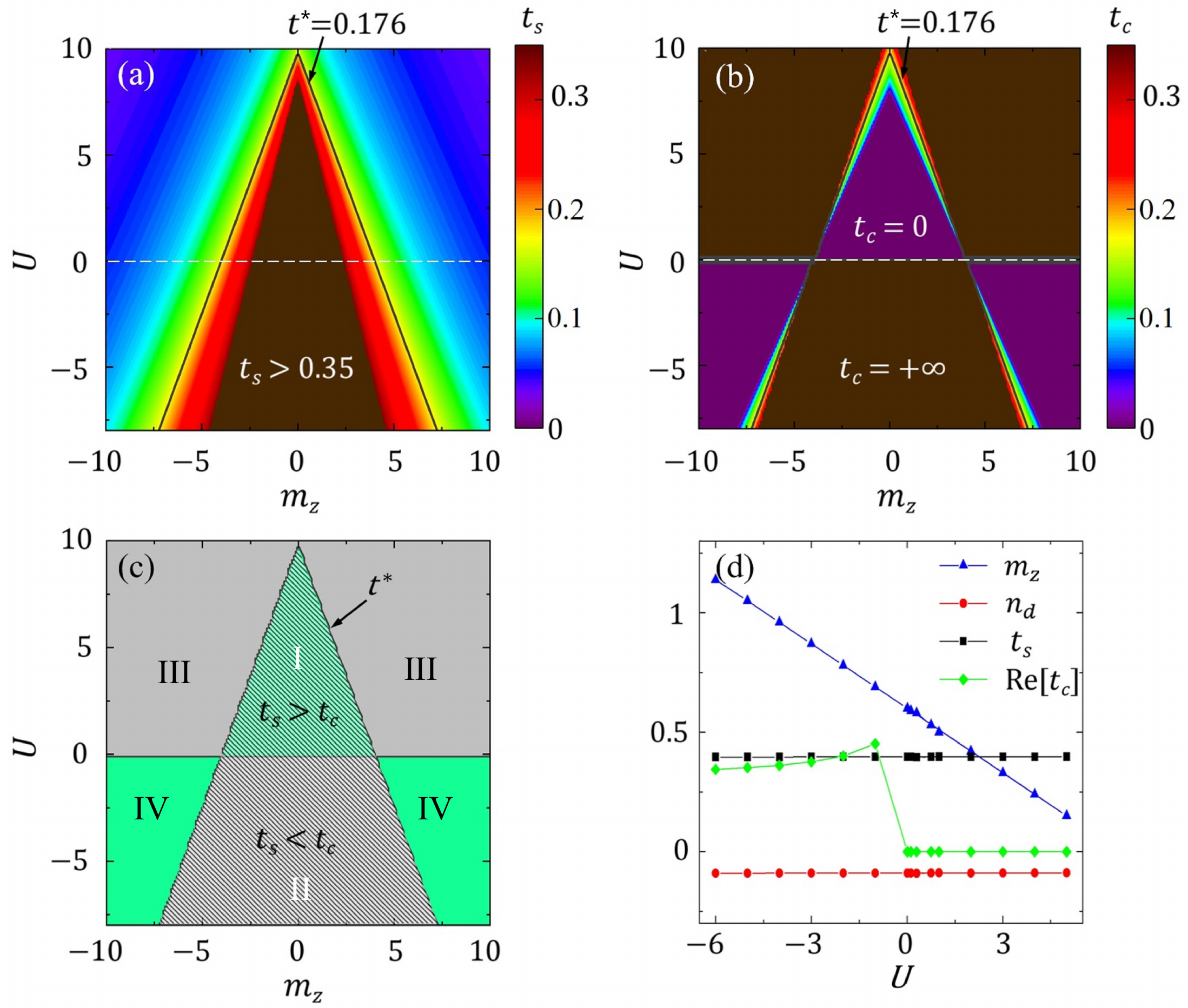}
\caption{(a)-(b) Analytical results of $t_s$ and $t_c$, where $t^*=0.176$ is on the topological boundaries (black solid lines). The dark-brown regions present $t_s>0.35$ in (a) and $t_c=+\infty$ in (b) respectively, while the dark-purple regions in (b) give $t_c=0$. These regions actually correspond to the complex values of $t_s$ and $t_c$ since a large truncation $\mathcal{O}(U^3n_d^3)$ is taken. After reducing the truncation error, $t_s$ gives the real value but $t_c$ still gives complex value due to no solution. (c) Sign of $t_s-t_c$, where the shaded area is the equilibrium topological phase region. (d) Analytical results of $t_s$, $n_d$ and $\text{Re}[t_c]$ at $n_d\approx -0.09$, showing the linear scaling of $t_s$ and $n_d$. Here we set $t_{\text{so}}=t_0=1$.}
\label{Fig3}
\end{figure}

Besides, it should be emphasized that here the time-dependent particle density only becomes self-consistent, while the instantaneous wave function is not, which has subtle difference from the equilibrium mean-field Hamiltonian where both particle density and wave function are completely self-consistent [see Appendix \ref{appendix-3}]. This implies that the different instantaneous states may lead to the same value of $n_{\uparrow(\downarrow)}(t_s)$ and there may be multiple time points fulfilling the self-consistent condition. We hereby choose the smallest time point as $t_{s}$ due to the consideration of short-time evolution [see Fig.~\ref{Fig2}(a)]. Also, a finite $t_s$ shall definitely appear in the short-time evolution, since $n_{\uparrow(\downarrow)}(t)$ increases or decays in this regime and the state gradually reaches a steady state after $t>t_s$. 

The explicit form of $t_{s}$ is usually very complex, but we can obtain an analytical result in the short-time evolution for a weak interaction strength or a small $|Un_{d}|$. After some straightforward calculations, $t_s$ is given by [see Appendix~\ref{appendix-2}]
\begin{equation}\label{a-tn}
t_s\approx \sqrt{\frac{N_1}{N_0+\sqrt{N_2+N_3 (m_z-n_dU)^2}}}
\end{equation}
with $N_0=12t_{\text{so}}^2$, $N_1=3(1\pm 2n_d)$, $N_2=-28t_{\text{so}}^4(\pm 10n_d-1)-72t_0^2t_{\text{so}}^2(\pm 2n_d+1)$, and $N_3=-28t_{\text{so}}^2(\pm 2n_d+ 1)$, where $\pm$ corresponds to the sign of $m_z$. We clearly observe that $t_{s}$ is associated with $n_{d}$ and $t_\text{so}$. Since the contour lines of $n_{d}$ have the linear form of interaction strength $U$ and Zeeman coupling $m_{z}$, this characteristic time $t_{s}$ also has a similar linear scaling for a fixed $t_\text{so}$, which is presented in Fig.~\ref{Fig3}(a) and fully matches with the numerical results.

\subsection{Dynamical topological phase transition and characteristic time $t_c$}

We now theoretically determine the characteristic time $t_{c}$ from the dynamical topological phase transition of post-quenched Hamiltonian $\mathcal{H}_{\mathbf{k}}(t)$. As we known, the dynamical topological phase transition is a phase transition in time driven by sharp internal changes in the properties of a quantum many-body state and not driven by an external control parameter~\cite{heyl2018dynamical,heyl2019dynamical}. Quantum quenching is one way to induce the dynamical phase transitions. When we performing the quench dynamics for this system, the value of $|n_{d}(t)|$ generally decays with the time evolution and approaches steady since the initial state is prepared into a nearly fully polarized state for $t<0$ and has $n_d(t=0)\approx\pm 0.5$. When the interaction strength satisfies $U>-8t_0$, we find that the time-dependent effective Zeeman coupling $m^{\text{eff}}_z(t)$ only passes through one phase transition point, i.e., $4t_{0}$ for $m_{z}>0$ or $-4t_{0}$ for $m_{z}<0$. Moreover, the corresponding crossover may occurs multiple times. Hence $\mathcal{H}_{\mathbf{k}}(t)$ only changes from the trivial (topological) regime at $t=0$ to the topological (trivial) regime for $t>0$, when the interaction is repulsive (attractive). To describe the critical time of the dynamical topological phase transition, the characteristic time $t_{c}$ is naturally defined as
\begin{equation}\label{eq:6}
t_{c}\equiv\text{min}\{t|m^{\text{eff}}_z(t)=\pm 4t_0\}.
\end{equation}
We have $t_c=+\infty$ and $0$ ($t_c=0$ and $+\infty$) in the repulsive (attractive) interaction regime when $|m^{\text{eff}}_z(t)|>4t_0$ and $|m^{\text{eff}}_z(t)|<4t_0$ respectively. Here $t_c=+\infty$ of the repulsive interaction means that the topology of $\mathcal{H}_\mathbf{k}(t)$ is always trivial in a long evolved time, while a finite $t_c$ implies that its topology is changed from the trivial case with $t<0$ to the topological case with $t=t_c$. 

With the precise definition of $t_c$, we next numerically show it in Fig.~\ref{Fig2}, where $m^{\text{eff}}_z(t)$ only passes though the phase transition point $m^{\text{eff}}_z(t_c)=4t_0$ and decays (increases) for the repulsive (attractive) interaction in the short time region. Note that this feature can be described by the analytical $n_d(t)$ in Appendix \ref{appendix-2}, where we have $\dot{m}_z^{\text{eff}}(t)\gtrless 0$ for $Um_z\lessgtr 0$ due to $\dot{m}_z^{\text{eff}}(t)=-U\dot{n}_d(t)$. Besides, we should choose the minimum time scale in the Eq.~\eqref{eq:6} which is similar to $t_s$, since the instantaneous state is not self-consistent and there may be multiple time points to satisfy this requirement.

Like Eq.~\eqref{a-tn}, we also obtain an analytical result for $t_c$ in the short-time evolution [see Appendix \ref{appendix-2}] as follows:
\begin{equation}\label{a-tp}
t_c\approx \sqrt{\frac{P_0-\sqrt{{P_2-3P_1{{(\pm m_z-4t_0)}/{U}}}}}{P_1}}
\end{equation}
with $P_0=6t_{\text{so}}^2$, $P_1=4t_{\text{so}}^2(5t_{\text{so}}^2+19t_0^2)$, and $P_2=6t_{\text{so}}^2(t_{\text{so}}^2-19t_0^2)$. Compared with $t_s$, the characteristic time $t_c$ presents a completely different scaling form in terms of the Zeeman coupling and interaction strength, expect for the topological phase boundaries~[see Fig.~\ref{Fig3}(b)]. This also implies that $t_s=t_c$ happens on these topological boundaries. On the other hand, we observe that the finite $t_c$ only appears around the topological phase boundaries. The reason is that $\mathcal{H}_{\mathbf{k}}(t)$ always remains topological or trivial case when it is far away from the phase boundaries~[This also causes the unchanged dynamical invariant $\nu (t)$ of Eq.~\eqref{eq:tcn} in the evolution], 
as we show in Figs.~\ref{Fig5}(b) and \ref{Fig5}(d).

\subsection{Characteristic time $t^*$ and its linear scaling}

{\color{black}We now theoretically determine the characteristic time $t^{*}$. Since the post-quenched parameters on the topological phase boundaries keep $m^{\text{eff}}_z(t_s)=m^{\text{eff}}_z(t_c)$, a special evolution time can be defined on these topological phase boundaries by
\begin{equation}\label{tx}
t^*\equiv\text{min}\{t|n_d(t)=n^*_d\},
\end{equation}
which holds the nontrivial relation $t_s=t_c$ [see Figs.~\ref{Fig3}(a) and~\ref{Fig3}(b)]. This point can be clearly observed by taking $m_z-Un_d=\pm 4 t_0$ in the Eqs.~\eqref{a-tn} and \eqref{a-tp}, which synchronously gives
\begin{equation}\label{tst}
t^{*}\approx\sqrt{\frac{B_1}{B_0+\sqrt{B_2}}},
\end{equation}
where $B_0=6t_\text{so}^2$, $B_1=3(1\pm 2n^{*}_d)/2$, and $B_2=-114(1\pm 2n^*_d)t_0^2t_\text{so}^2-6(\pm 10n^{*}_d-1)t_\text{so}^4$; see Appendix~\ref{appendix-2}. For the parameters on the topological boundaries, the gap closing occurs at Dirac points $\boldsymbol{\Gamma}$ or $\mathbf{M}$, which is naturally captured by $t^*$. Besides, it should be noted that $t^*$ is only a time scale corresponding to these topological boundaries and has slightly difference with the previous $t_s$ and $t_c$ which can cover all parameters. In particular, we see that the analytical $t^*$ is associated with $n^*_d$ and $t_\text{so}$, where $n^*_d$ is usually a constant in the topological phase boundaries and has the linear scaling for the Zeeman coupling and interaction. Hence $t^*$ shall inherit these linear properties and only depend on the spin-orbit coupled strength. As it is shown in Fig.~\ref{Fig3}, we have $t^*\propto 1/t_0$ for a strong spin-orbit coupling with $t_{\text{so}}=t_0$, which theoretically gives $t^*= 0.176$ when $t_{\text{so}}=t_0=1$ and this matches with the numerical result $t^*=0.180$ [see Fig.~\ref{Fig5}(b)].} 

\section{Mean-field phase diagram determined by time scales}\label{sec:Mean-field phase diagram determined by time scales}

In this section, we establish the dynamical characterization via the above three time scales. Further, we accurately determine the Chern number by examining the spin dynamics at four Dirac points. Particularly, we demonstrate the correctness of mean-field phase diagram by comparing the dynamical measurement results with the theoretical self-consistent results. 

\subsection{Dynamical characterization}
 
In the previous quench dynamics, this interacting system shall emerge the different time scales in the different post-quenched topological phases with the repulsive (attractive) interaction, when $\mathcal{H}_\mathbf{k}(t)$ is trivial (topological) at $t=0$. Specifically, the relations of three time scales are given via the four regimes of Fig.~\ref{Fig3}(c). Regime I ($U>0$, topological phase): $t_s>t_c$, $t^*>t_c$, and $t^*<t_s$; Regime II ($U<0$, topological phase): $t_s<t_c$, $t^*<t_c$, and $t^*<t_s$; Regime III ($U>0$, trivial phase): $t_s<t_c$, $t^*<t_c$, and $t^*>t_s$; Regime IV ($U<0$, trivial phase): $t_s>t_c$, $t^*>t_c$, and $t^*>t_s$. Equivalently, the correspondence between the characteristic times and the equilibrium topological properties is given by
\begin{equation}\label{eq:7}
|\text{Ch}_1|=\left\{
\begin{aligned}
1 & , & \text{for}
\begin{cases}
        t_s>t^*>t_c,~U>0,\\
        t^*<t_s<t_c,~U<0,
\end{cases}\\
0 & , & \text{for}
\begin{cases}
        t_s<t^*<t_c,~U>0,\\
        t^*>t_s>t_c,~U<0.
\end{cases}\\
\end{aligned}
\right.
\end{equation}
The formula provides a convenient dynamical way to determine the mean-field topological phase by comparing any two characteristic time scales. Moreover, both the numerical results [Fig.~\ref{Fig2}] and the analytical results [Fig.~\ref{Fig3}(c)] confirm this way, especially for weak interaction $U$ or small $|Un_d|$, as we show in Fig.~\ref{Fig3}(d). Hence we only need to prepare an initial nearly fully polarized state and perform quenching, and then the mean-field topological phases can be directly determined by comparing any two time scales. 

\begin{figure}[t]
\includegraphics[width=\columnwidth]{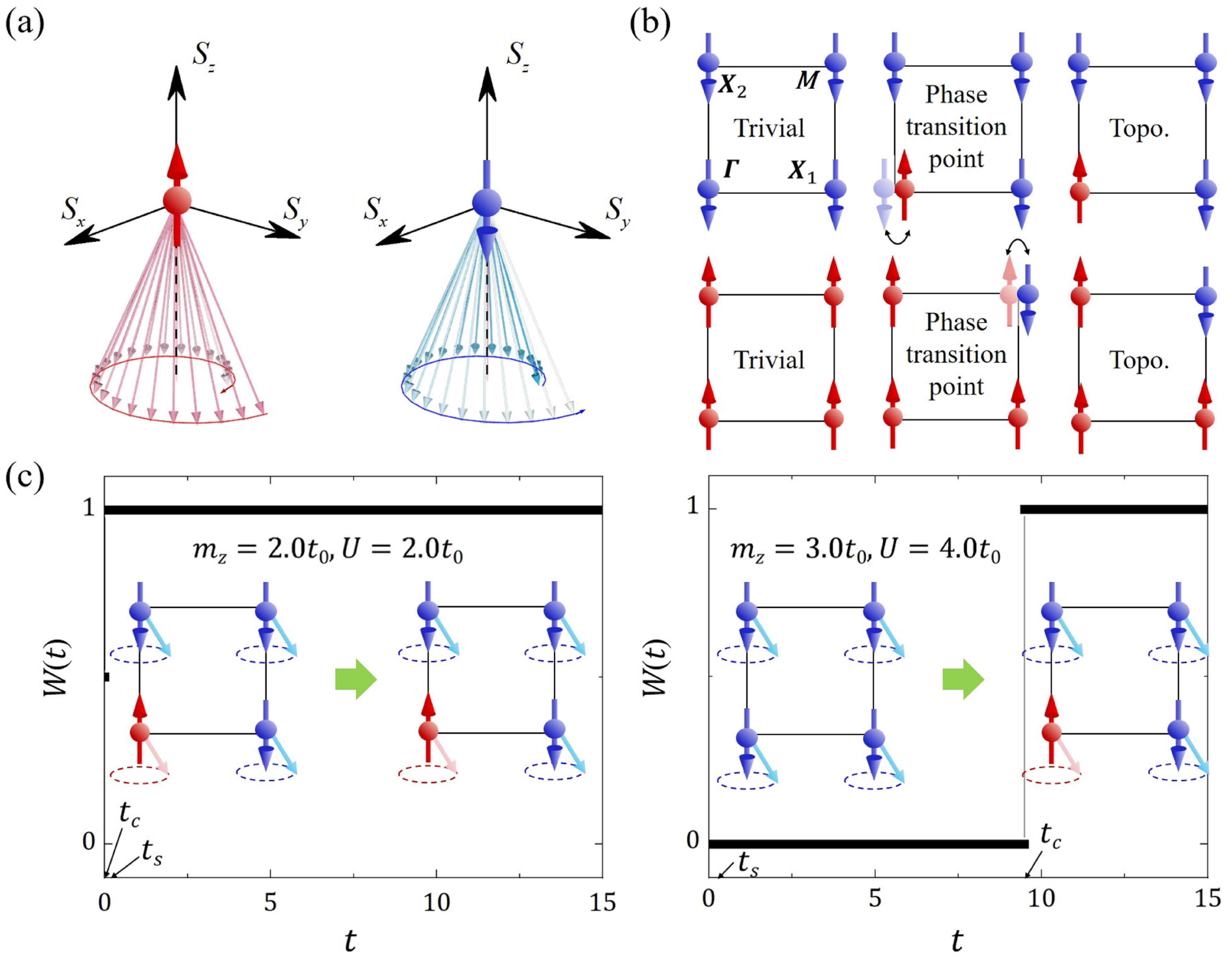}
\caption{(a) Schematic diagram for the motion of the nearly fully polarized spin at Dirac points $\mathbf{D}_i$. The clockwise (counterclockwise) motion of the spin implies the upward (downward) polarization of Hamiltonian at these points. (b) The polarization directions at Dirac points $\mathbf{D}_i$ determine the topology of post-quenched Hamiltonian. (c) Topological number $W(t)$ for $\mathcal{H}_{\mathbf{k}}(t)$. We have $\text{Ch}_1=1$ for the phase with $(m_z,U)=(2,2)$ due to $t_{c}=0<t_{s}=0.165$, while $\text{Ch}_1=0$ for the phase with $(m_z,U)=(3,4)$ due to $t_{c}=9.485>t_{s}=0.230$. The insets show the changes of polarization directions at Dirac points $\mathbf{D}_i$ for $\mathcal{H}_{\mathbf{k}}(t)$. Here we set $t_{\text{so}}=t_0=1$, $m^{(c)}_{x}=m_z$, $m^{(c)}_y=0$, and $m^{(c)}_z=100$.
}
\label{Fig4}
\end{figure}

\subsection{Determination of Chern number}

For the equilibrium mean-field topological phases of Eq.~\eqref{eq:7}, we further need to accurately determine the Chern number by considering the features of spin dynamics. Specifically, the time-dependent spin evolution of this interacting system is described by the modified Landau-Lifshitz equation~\cite{gilbert2004phenomenological}
\begin{equation}
\dot{\mathbf{S}}(\mathbf{k},t)=\mathbf{S}(\mathbf{k},t)\times 2\mathbf{h}(\mathbf{k},t)
\end{equation}
with 
\begin{equation}
\begin{split}
& S_x(\mathbf{k},t)=[\chi_\mathbf{k}(t)\eta^{*}_\mathbf{k}(t)+\chi^{*}_\mathbf{k}(t)\eta_\mathbf{k}(t)]/2,\\
& S_y(\mathbf{k},t)=i[\chi_\mathbf{k}(t)\eta^{*}_\mathbf{k}(t)-\chi^{*}_\mathbf{k}(t)\eta_\mathbf{k}(t)]/2,\\
& S_z(\mathbf{k},t)=[|\chi_\mathbf{k}(t)|^2-|\eta_\mathbf{k}(t)|^2]/2.\\ 
 \end{split}
\end{equation}
The topological number $W(t)$ of the time-dependent post-quenched Hamiltonian $\mathcal{H}_{\mathbf{k}}(t)$ is determined by the spin dynamics of $S_z(\mathbf{D}_i,t)$ at four Dirac points $\mathbf{D}_i$. The Chern number $\text{Ch}_1$ of the equilibrium mean-field Hamiltonian is then obtained by $W(t_s)$.

We next show two fundamental spin dynamical properties of this interacting system. Firstly, the spin at four Dirac points of the nearly fully polarized initial state will move around the corresponding polarization direction of $\mathcal{H}_{\mathbf{k}}(t)$ after the quantum quench. When the polarization direction is reversed, e.g., from upward to downward, the motion of spin will be reversed at the same time, e.g., from clockwise to counterclockwise; see Fig.~\ref{Fig4}(a). For this, we can use the motion of spins to determine the polarization direction of $\mathcal{H}_{\mathbf{k}}(t)$ at Dirac points. Second, since the polarization direction is associated with the parity eigenvalue of the occupied eigenstate, the topology of $\mathcal{H}_{\mathbf{k}}(t)$ can be identified from the polarization directions at four Dirac points~\cite{liu2013detecting}, as shown in Fig.~\ref{Fig4}(b). For instance, the polarization directions at four Dirac points are the same for an initial trivial Hamiltonian $\mathcal{H}_{\mathbf{k}}(t)$. Once it enters the topological regime, the motion of spin at $\boldsymbol{\Gamma}$ or $\mathbf{M}$ point will be reversed, manifesting the change of the corresponding polarization direction. This signal can be captured by $\text{sgn}[\dot{S}_z(\mathbf{D}_i,t)]$, as shown in Fig.~\ref{Fig5}(d). Correspondingly, the topological transition time of the motion of spin just gives the characteristic time $t_c$.

With above two fundamental properties, we can define the time-dependent dynamical invariant for $\mathcal{H}_{\mathbf{k}}(t)$ as
\begin{equation}\label{nut}
(-1)^{\nu(t)}=\prod_{i}\text{sgn}[\dot{S}_z(\mathbf{D}_i,t)],
\end{equation}
with $\nu(t)=1$ for the topological regime $|m^{\text{eff}}_z(t)|<4t_0$ and $\nu(t)=0$ for the trivial regime $|m^{\text{eff}}_z(t)|>4t_0$, respectively. {\color{black}The topological transition time of this dynamical invariant gives the characteristic time $t_c$ as follows:
\begin{equation}\label{eq:tcn}
t_c=\left\{
\begin{aligned}
0,~~& \text{for}~~\nu (t)=1~(0),U>0~(U<0),\\
\text{finite},~~& \text{for}~~\nu (t)=1\leftrightarrow \nu (t)=0,\\
+\infty ,~~& \text{for}~~\nu (t)=0~(1),U>0~(U<0). \\
\end{aligned}
\right.
\end{equation}
Here a finite $t_c$ is captured by the critical time of changing of $\nu (t)$ from $1$ ($0$) to $0$ ($1$) [see Figs.~\ref{Fig3}(b) and \ref{Fig5}(d)], while the unchanged $\nu (t)$ in a long-time evolution hosts $t_c=0$ or $+\infty$.} Now the topological number $W(t)$ of $\mathcal{H}_\mathbf{k}(t)$ and the Chern number of $\mathcal{H}^{\text{MF}}_{\mathbf{k}}$ can be exactly given by
\begin{equation}\label{ch}
W(t)=\frac{\nu(t)}{2}\sum_{i}\text{sgn}[\dot{S}_z(\mathbf{D}_i,t)],~~\text{Ch}_1=W(t_s).
\end{equation}
As an example, the numerical results in Fig.~\ref{Fig4}(c) show the nontrivial topology with $\text{Ch}_1=1$ and the trivial topology with $\text{Ch}_1=0$ for the post-quenched system with $(m_z,U)=(2,2)$ and $(3,4)$, respectively. On the other hand, since there is only one topological phase transition point in the dynamical evolution, $W(t)$ has the same value at $t_s$ with $t^*$. Hence we also have
\begin{equation}\label{ch_app}
\text{Ch}_1\simeq W(t^*).
\end{equation}
It should be emphasized that any one formula of the Eqs.~\eqref{ch} and \eqref{ch_app} facilitates the determination of Chern number of the equilibrium mean-field topological phases for the post-quenched system.  
In the following, we take Eq.~\eqref{ch_app} in the experiment for identifying the Chern number due to the easy measurement of $t^*$.

\section{Experimental detection}\label{sec:Experimental detection}

{\color{black}Based on above topological characterization, we next propose a feasible experimental scheme to detect the mean-field topological phase diagram by directly measuring the two characteristic time scales $t_c$ and $t^*$, without any prior post-quenched parameters such as $m_z$ and $U$. Note that the measurement of $t_{s}$ is almost impossible in the recent experiment, since it does not have any measurable quantities but only corresponds to the self-consistency of post-quenched Hamiltonian. 

\begin{figure}[!tbp]
\includegraphics[width=\columnwidth]{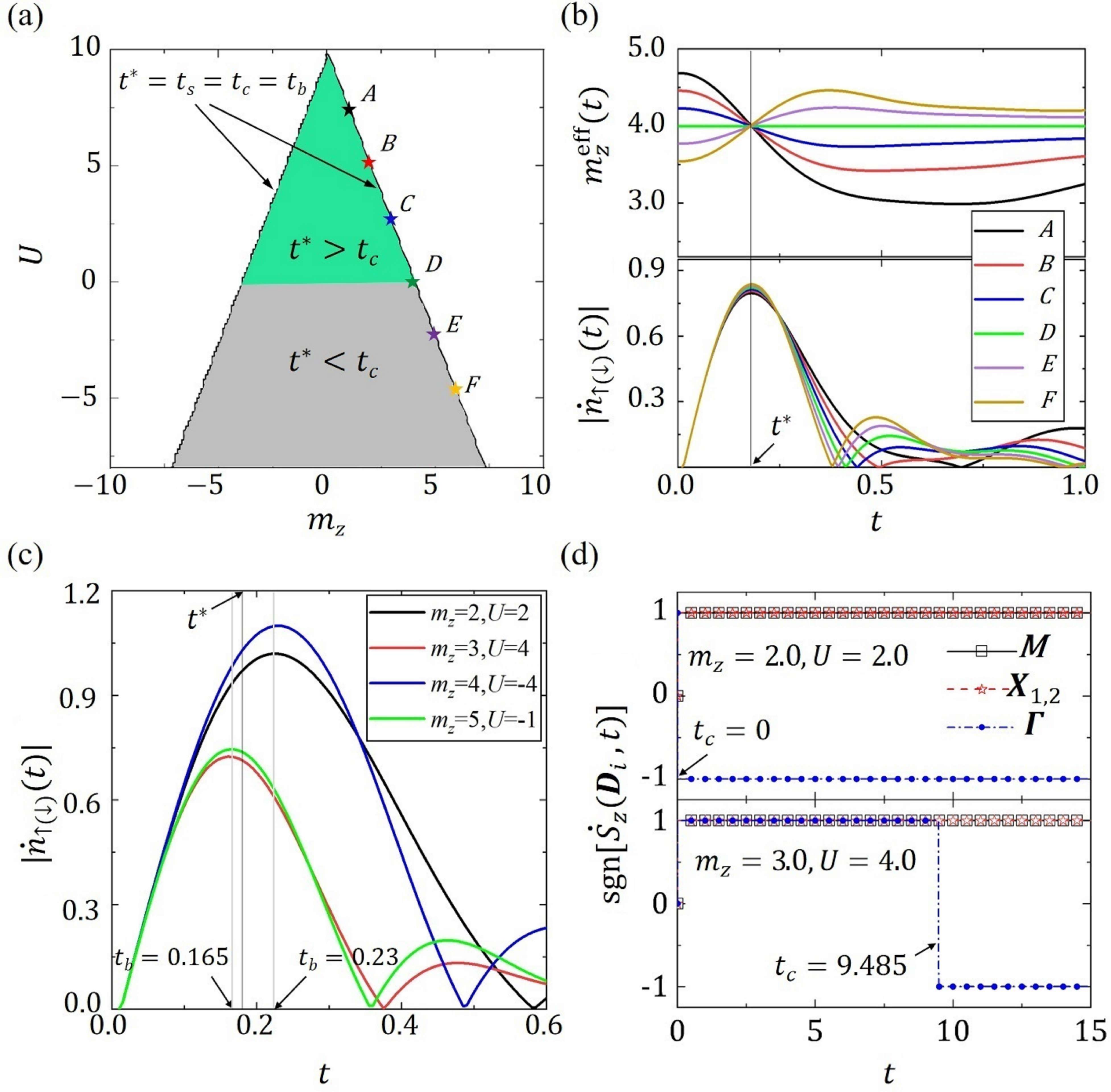}
\caption{(a) Mean-field phase diagram determined by the characteristic times $t_c$ and $t^*$, where the post-quenched parameters $A$-$F$ (stars) are chosen on the topological phase boundaries with $(m_z,U)=(1.0,7.38)$, $(2.0,4.92)$, $(3.0,2.46)$, $(4.0,0)$, $(5.0,-2.46)$, and $(6.0,-4.92)$. (b) Dynamical evolution of $m^{\text{eff}}_z(t)$ and $|\dot{n}_{\uparrow(\downarrow)}(t)|$ for $A$-$F$, where $t_c=t^{*}=0.180$ (upper) and $t_b=t^{*}=0.180$ (lower) with the same scaling. (c) Dynamical evolution of $|\dot{n}_{\uparrow(\downarrow)}(t)|$, where $t_b=0.165$ for $(m_z,U)=(3.0,4.0)$ and $(5.0,-1.0)$ while $t_b=0.230$ for $(m_z,U)=(2.0,2.0)$ and $(4.0,-4.0)$.
(d) Sign of $\dot{S}_z(\mathbf{D}_i,t)$ at four Dirac points. We have $\text{Ch}_1=1$ for the topological phase with $(m_z,U)=(2.0,2.0)$ due to $t_c=0<t^*$ and $\text{Ch}_1=0$ for the trivial phase with $(m_z,U)=(3.0,4.0)$ due to $t_c=9.485>t^*$. Here we set $t_{\text{so}}=t_0=1$, $m^{(c)}_{x}=m_z$, $m^{(c)}_y=0$, and $m^{(c)}_z=100$.
}
\label{Fig5}
\end{figure}

In a realistic experiment, identifying the time scale of $t_c$ needs to measure the spin dynamics of $S_z(\mathbf{D}_i,t)$ at four Dirac points $\mathbf{D}_i$ in a short-time evolution and calculate their changing with the evolved time, i.e., $\dot{S}_z(\mathbf{D}_i,t)$. And then, $t_c$ is captured by the topological transition time of $\nu (t)$ based on the Eq.~\eqref{eq:tcn}. On the other hand, we need to measure the time-dependent particle density $n_{\uparrow(\downarrow)}(t)$ and calculate the fastest changing of $n_{\uparrow(\downarrow)}(t)$ with the evolved time and identify the time scale of $t^*$ by an auxiliary time
\begin{equation}\label{eq:7s}
t_b\equiv \text{min}\{t|\ddot{n}_{\uparrow(\downarrow)}(t)=0\}.
\end{equation}
We see that $t_b$ characterizes the evolution of $n_{\uparrow(\downarrow)}(t)$ changes fastestly, i.e., $|\dot{n}_{\uparrow(\downarrow)}(t)|$ is maximal. Particularly, $t_{b}=t^*$ occurs on the topological phase boundaries although it is larger (smaller) than $t^{*}$ in the topological (trivial) regime [see Fig.~\ref{Fig5}(c)]. This point can be clearly observed by an analytical result from Eq.~\eqref{eq:7s}, i.e.,
\begin{equation}\label{eq:t*}
t^*\approx \sqrt{\frac{3}{Q_0+Q_1}}
\end{equation}
with $Q_0=\sqrt{-135t_0^2+606t_0^2t_\text{so}^2+57t_\text{so}^4}$ and $Q_1=57t_0^2+15t_\text{so}^2$. This formula is consistent with Eq.~(\ref{tst}). Also, we have $t^*\approx 0.178$ for $t_0=t_\text{so}=1$, which is approximately equal to $t^*=0.176$ in the Eq.~(\ref{tst}). 

With above observation, the mean-field topological phase diagram can be obtained by only comparing both time scales of $t_c$ and $t^*$; see Fig.~\ref{Fig5}(a). Considering that the 2D QAH model has realized in the cold atoms~\cite{wu2016realization} and the interaction can be can be tuned in
experiment~\cite{bloch2008many}, we hereby provide the following four concrete steps to detect the equilibrium mean-field topological phases for the cold atom experiment: 
\begin{enumerate}[  ]
\item \emph{Step I}: preparing the nearly fully polarized initial state for the system with $U=0$, which can be reached by tuning say magnetic field for ultracold atoms. By further varying the two-photon detuning via the bias magnetic field to produce a large $m_z$, together with turning off one of the electro-optic modulators to generate the small constant magnetization $m_x\sigma_x$ or $m_y\sigma_y$~\cite{zhang2019dynamical}, the nearly fully polarized state is now obtained;
\item\emph{Step II}: performing quench dynamics for this system with $U=0$. We suddenly change the magnetic field to produce a finite $m_z=\pm 4t_0$ and turn off both the electro-optic modulators to generate $m_{x(y)}=0$, which drives the system evolve with the time. And then, we measure the time-dependent particle density $n_{\uparrow(\downarrow)}(t)$. The time scale of $t^*$ is obtained by the fastest changing time point of $n_{\uparrow(\downarrow)}(t)$. Note that this $t^*$ has the linear scaling on the mean-field topological phase boundaries [see Fig.~\ref{Fig3}(c)] and is independent of $U$. Hence $t^*$ can be directly employed to the cases of $U\neq 0$;
\item\emph{Step III}: preparing the nearly fully polarized initial state for the system with $U\neq 0$. This step is similar to Step I. Here $U$ can be an unknown value; 
\item\emph{Step IV}: performing quench dynamics for this system with the nonzero $U$. For this we suddenly tune the system to a regime with an unknown suitable $m_z$ and turn off both the electro-optic modulators to generate $m_{x(y)}=0$. Simultaneously, we measure the evolution of $S_z(\mathbf{D}_i,t)$ at four Dirac points and obtain their changing with the evolved time, i.e., $\dot{S}_z(\mathbf{D}_i,t)$. Then $t_c$ is captured by the topological transition time of $\nu (t)$ based on the Eq.~\eqref{eq:tcn}. Finally, the mean-field topological phases are identified by comparing $t^*$ in the Step II and $t_c$ in the Step IV, where the Chern number is given by  $\text{Ch}_1\simeq W(t^*)$; see Eq.~\eqref{ch_app}. 
\end{enumerate}
With this scheme, the mean-field phase diagram of the interacting Chern insulator can be completely determined, which paves the way to experimentally study topological phases in interacting systems and discover new phases.}


\section{Conclusion and discussion}\label{sec:Conclusion}

In conclusion, we have studied the 2D QAH model with a weak-to-intermediate Hubbard interaction by performing quantum quenches. A nonequilibrium detection scheme for the mean-field topological phase diagram is proposed by observing three characteristic times $t_{s}$, $t_{c}$, and $t^*$ emerged in dynamics. We reveals three nontrivial dynamical properties: (i) $t_s$ and $t_c$ capture the emergence of dynamical self-consistent particle density and dynamical topological phase transition for the time-dependent post-quenched Hamiltonian respectively, while $t^*$ gives a linear scaling time on the topological phase boundaries; (ii) After quenching the Zeeman coupling from the trivial regime to the topological regime, $t_s>t^*>t_c$ ($t^*<t_s<t_c$) is observed for the repulsive (attractive) interaction; (iii) The Chern number of post-quenched mean-field topological phase can be determined by comparing any two time scales. These results provide a feasible scheme to detect the mean-field topological phases of an interacting Chern insulator and may be applied to the quantum simulation experiments. 

The present dynamical scheme can be applied to the Chern insulator with the weak-to-intermediate interaction. On the other hand, when the general strong interactions are considered, the system may host more abundant magnetic phases~\cite{ziegler2020correlated,ziegler2022large,tirrito2022large}. Generalizing these dynamical properties to the related studies and further identifying the rich topological phases would be an interesting and worthwhile work in the future.

\section*{ACKNOWLEDGEMENT}
We acknowledge the valuable discussions with Sen Niu and Ting Fung Jeffrey Poon. This work was supported by National Natural Science Foundation of China (Grants No. 11825401 and No. 11921005), the National Key R\&D Program of China (Project No. 2021YFA1400900), Strategic Priority Research Program of the Chinese Academy of Science (Grant No. XDB28000000), and by the Open Project of Shenzhen Institute of Quantum Science and Engineering (Grant No.SIQSE202003). Long Zhang also acknowledges support from the startup grant of the Huazhong University of Science and Technology (Grant No. 3004012191). Lin Zhang also acknowledges support from Agencia Estatal de Investigación (the R\&D project CEX2019-000910-S, funded by MCIN/AEI/10.13039/501100011033, Plan National FIDEUA PID2019-106901GB-I00, FPI), Fundaci{\'o} Privada Cellex, Fundaci{\'o} Mir-Puig, Generalitat de Catalunya (AGAUR Grant No. 2017 SGR 1341, CERCA program), EU Horizon 2020 FET-OPEN OPTOlogic
(Grant No. 899794), and Junior Leaders fellowship LCF/BQ/PI19/11690013 of "La Caixa” Foundation (ID100010434).

\appendix

\begin{figure}[!bp]
\includegraphics[width=\columnwidth]{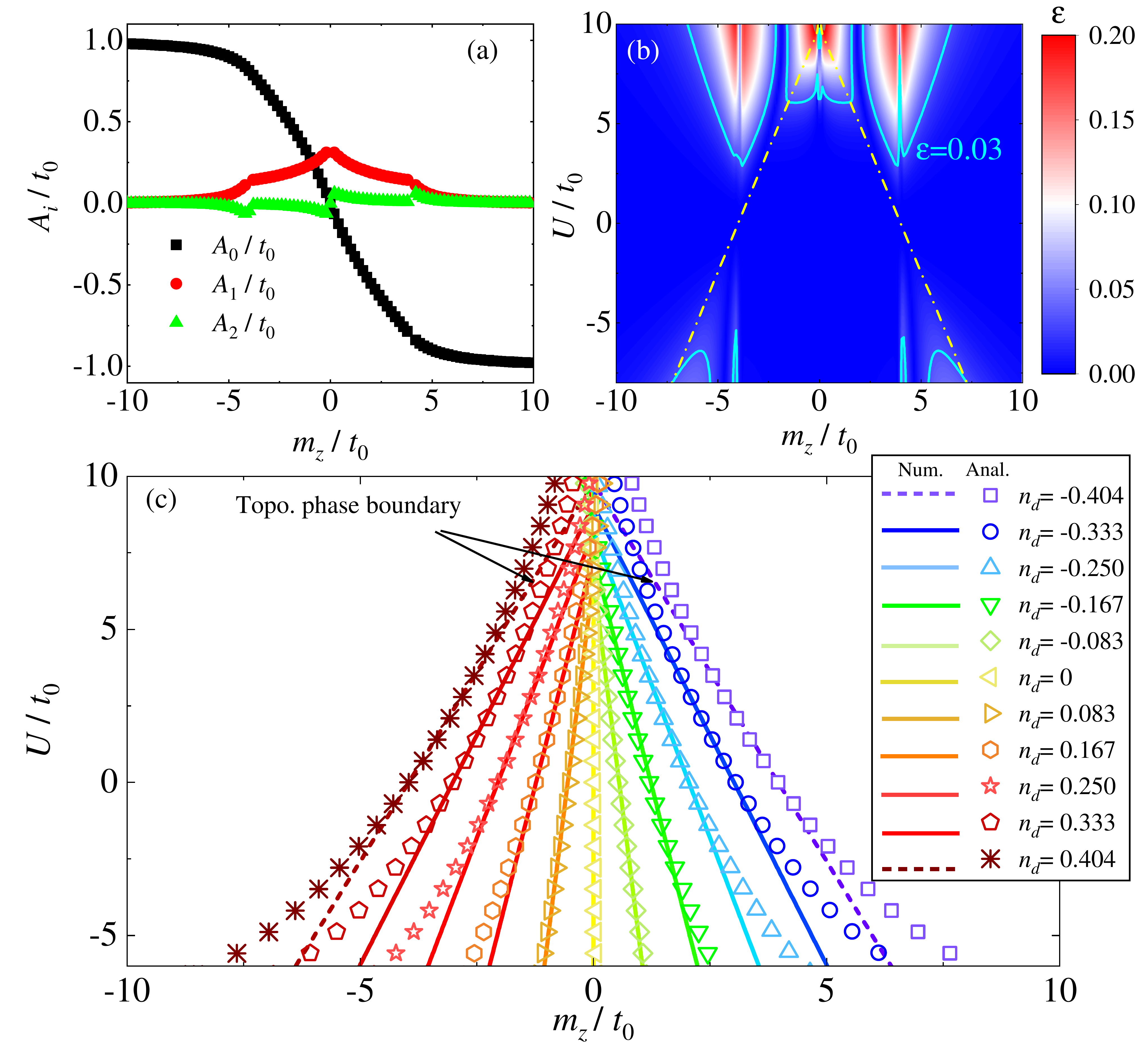}
\caption{(a) Analytical coefficients $A_0$, $A_1$, and $A_2$. (b) Absolute error $\varepsilon$ between the analytical results and the numerical results for $n_d$.
(c) Analytical results of $n_d$ corresponding to the numerical results shown in Fig.~\ref{Fig1}(d), giving the linear scaling. Here we set $t_{\text{so}}=t_0$.
}
\label{Fig-a1}
\end{figure}

\section{Equilibrium analytical results}\label{appendix-1}

In presence of the non-zero interaction, the equilibrium mean-field Hamiltonian $\mathcal{H}^{\text{MF}}_\mathbf{k}$ self-consistently gives the difference of density for spin-up and spin-down particles as $n_d\equiv (n_\uparrow-n_\downarrow)/2$ under the mean field theory. Here we explicitly show 
\begin{equation}\label{ap-2}
n_d=\frac{1}{N}\sum_{\mathbf{k}}\frac{-h_{z,\mathbf{k}}+Un_d}{2\sqrt{h_{x,\mathbf{k}}^2+h_{y,\mathbf{k}}^2+(h_{z,\mathbf{k}}-Un_d)^2}}.
\end{equation}
Considering that $|n_d|<0.5$ is a small value due to the total particle density is conserved, i.e., $n_{\uparrow}+n_{\downarrow}=1$, we use Taylor series at $n_d=0$ for the right-hand term of Eq.~(\ref{ap-2}), which gives $A_0+A_1Un_d+A_2U^2n_d^2+\cdots +A_{l}U^{l}n_d^{l}$ with $(l+1)$th-order truncation $\mathcal{O}(U^{l+1}n_d^{l+1})$. Here the coefficients $A_{i=0,1,\cdots,l}$ satisfy $\partial A_0/\partial h_{z,\mathbf{k}}=-A_1,\partial A_1/\partial h_{z,\mathbf{k}}=-2A_2,\partial A_2/\partial h_{z,\mathbf{k}}=-3A_3,\cdots$. In principle, we can obtain $n_d$ by solving $n_d\approx A_0+A_1Un_d+A_2U^2n_d^2+\cdots+A_{l}U^{l}n_d^{l}$. The analytical $n_d$ for $l=2$ at a weak interaction or a small Zeeman coupling is approximately given by 
\begin{equation}\label{a-nd}
n_d\approx \frac{2-A_1U-\sqrt{(2-A_1U)^2-4A_0A_2U^2}}{2A_2U^2},
\end{equation}
with $A_0=\frac{1}{N}\sum_{\mathbf{k}}\frac{-h_{z,\mathbf{k}}}{e_{\mathbf{k}}}$, $A_1=\frac{1}{N}\sum_{\mathbf{k}}\frac{h_{x,\mathbf{k}}^2+h_{y,\mathbf{k}}^2}{e_{\mathbf{k}}^3}$, and $A_2=\frac{1}{N}\sum_{\mathbf{k}}\frac{3(h_{x,\mathbf{k}}^2+h_{y,\mathbf{k}}^2)h_z}{2e_{\mathbf{k}}^5}$. The Eq.~(\ref{a-nd}) shows the scaling property of $n_d$ for $m_z$ and $U$.

Besides, we observe that $A_i$ can converge to zero for an increased $i$; see Fig.~\ref{Fig-a1}(a). For a 3rd-order truncation $\mathcal{O}(U^{3}n_d^{3})$ in Eq.~(\ref{a-nd}), the absolute error $\varepsilon$ between analytical and numerical results presents $\varepsilon<0.03$ for $|U|<6t_0$, which implies that the precision of 3rd-order truncation is enough. Therefore, this Eq.~(\ref{a-nd}) is applicable to the almost completely topological phase region and all the topological phase boundaries and a large trivial regions, as shown in Fig.~\ref{Fig-a1}(b). Also, the analytical results are almost confirmed with the numerical results for both a weak interaction or a small $m_z$, as shown in Fig.~\ref{Fig-a1}(c).

\begin{figure}[t]
\includegraphics[width=\columnwidth]{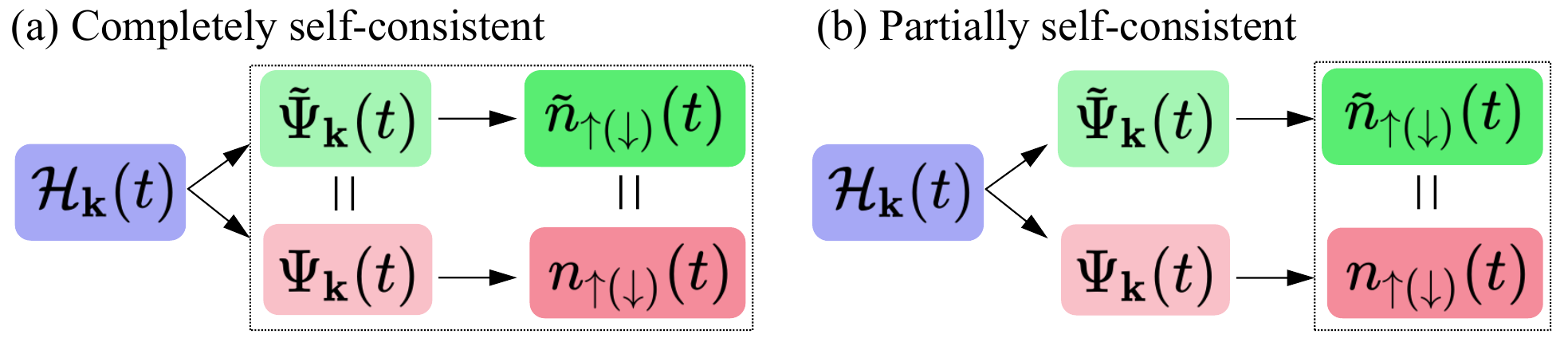}
\caption{Flowcharts for the self-consistent calculation of particle density. (a) Completely self-consistent. (b) Partially self-consistent.}
\label{Fig-a2}
\end{figure}

\section{Nonequilibrium analytical results}\label{appendix-2}

Considering that the spin of initial state is nearly fully polarized to downward or upward, we can take $[\chi_{\mathbf{k}}(0),\eta_{\mathbf{k}}(0)]^T\approx(0,1)^T$ or $[\chi_{\mathbf{k}}(0),\eta_{\mathbf{k}}(0)]^T\approx(1,0)^T$, which corresponds to the post-quenched $m_z>0$ or $m_z<0$. Further, the change of $n_d(t)$ is small during a short-time evolution and it can be regarded as a constant in a very short time. By solving Eq.~(\ref{eq:2}), we can obtain $n_d(t)\approx n_{d,t}$ and which is given by
\begin{equation}\label{evolution}
 n_{d,t}=\pm\frac{1}{N}\sum_{\mathbf{k}}\frac{-e_{\mathbf{k},t}^2+(h_{x,\mathbf{k}}^2+h_{y,\mathbf{k}}^2)\left(1-\cos2te_{\mathbf{k},t}\right)}{2e_{\mathbf{k},t}^2},
\end{equation}
where $e_{\mathbf{k},t}=\sqrt{h_{x,\mathbf{k}}^2+h_{y,\mathbf{k}}^2+(h_{z,\mathbf{k}}-n_{d,t}U)^2}$ and $\pm$ corresponds to the sign of the post-quenched $m_z$. Similarly, we use Taylor series at $n_{d,t}=0$ for the right-hand term of Eq.~(\ref{evolution}), which gives $D_0+D_1Un_{d,t}+D_2U^2n_{d,t}^2+\cdots +D_{l}U^{l}n_{d,t}^{l}$ with $(l+1)$th-order truncation $\mathcal{O}(U^{l+1}n_{d,t}^{l+1})$.

We next take the 3rd-order truncation $\mathcal{O}(U^{3}n_{d,t}^{3})$ to approximately obtain $n_{d,t}$ as follows:
\begin{equation}\label{dy approximate analytical}
n_{d,t}\approx \frac{2-D_1U-\sqrt{(2-D_1U)^2-4D_0D_2U^2}}{2D_2U^2}
\end{equation}
with $D_0=\pm8 t_{\text{so}}^2t^2-[\pm 1\pm {8}t_{\text{so}}^2(m_z^2+3t_0^2+5t_\text{so}^2)t^4/3]$, $D_1=\pm{16}m_zt_\text{so}^2t^4/3$, and $D_2=-(\pm{8}t_\text{so}^2t^4/3)$. Note that here we have used Taylor series at $t=0$ for $D_{0,1,2}$ and have taken 6th-order truncation $\mathcal{O}(t^6)$ for $t$,. Then this approximate $n_{d,t}$ is only suitable for a short-time evolution or a small value of $|Un_{d,t}|$. Finally, we explicitly give $t_s$, $t_c$, $t^*$, and $t_b$ by solving the equations of $n_{d,t}=n_d$, $m_z-Un_{d,t}=\pm 4t_0$, $n_d(t)=n^*_d$, and $\ddot{n}_{d,t}=0$ respectively, where the theoretical results are shown in Eq.~(\ref{a-tn}), Eq.~(\ref{a-tp}), Eq.~(\ref{tst}), and Eq.~(\ref{eq:t*}).

\section{Dynamical self-consistent processes of particle density}\label{appendix-3}

There are two cases of the dynamical self-consistent processes to obtain the self-consistent particle density $\tilde{n}_{\uparrow(\downarrow)}(t)$. The first one is a completely self-consistent process. Namely, both wave function and particle density are self-consistently, as it is shown in Fig.~\ref{Fig-a2}(a). By diagonalizing the time-dependent Hamiltonian $\mathcal{H}_\mathbf{k}(t)$ and obtaining its eigenvector $\tilde{\Psi}_{\mathbf{k}}(t) =[\tilde{\chi}_{\mathbf{k}}(t),\tilde{\eta}_{\mathbf{k}}(t)]^T$ with negative energy. Next the instantaneous wave function $\Psi_{\mathbf{k}}(t)$ should equal to $\tilde{\Psi}_{\mathbf{k}}(t)$ at each momentum $\mathbf{k}$, i.e., $\Psi_{\mathbf{k}}(t)=\tilde{\Psi}_{\mathbf{k}}(t)$. It is clear that the self-consistent wave function directly gives $n_{\uparrow(\downarrow)}(t)=\tilde{n}_{\uparrow(\downarrow)}(t)$ from the Eq.~\eqref{eq:3} and Eq.~\eqref{eq:4}. Particularly, we emphasize that this system shall reach a dynamical balance and both $m^{\text{eff}}_z$ and $n_{\uparrow(\downarrow)}(t)$ do not evolve with the time, when $\Psi_{\mathbf{k}}(t)$ becomes the eigenvector of $\mathcal{H}_\mathbf{k}(t)$.

The second one is a partially self-consistent process. Namely, only particle density is self-consistently, as it is shown in Fig.~\ref{Fig-a2}(b). After obtaining the eigenvector $\tilde{\Psi}_{\mathbf{k}}(t) =[\tilde{\chi}_{\mathbf{k}}(t),\tilde{\eta}_{\mathbf{k}}(t)]^T$ with negative energy of $\mathcal{H}_\mathbf{k}(t)$, the difference with the above process is to directly take this $n_{\uparrow(\downarrow)}(t)$ equal $\tilde{n}_{\uparrow(\downarrow)}(t)$, of which we do not care the instantaneous states. It is clear that the self-consistent particle density are also given by $n_{\uparrow(\downarrow)}(t)=\tilde{n}_{\uparrow(\downarrow)}(t)$, but now the wave function is not self-consistent, i.e., $\Psi_{\mathbf{k}}(t)\neq\tilde{\Psi}_{\mathbf{k}}(t)$. Physically, the particle density is associated with the summation of the wave function at all $\mathbf{k}$, which implies that the wave function of each $\mathbf{k}$ may not be self-consistent but the summation for all $\mathbf{k}$ can be self-consistent. 

Both two processes can give the self-consistent particle density of post-quenched Hamiltonian $\mathcal{H}_\mathbf{k}(t)$, while the former is difficult to realize the completely self-consistent in a long-term evolution. The latter is easily obtained in a short-time evolution and is more helpful to capture the nontrivial dynamics.
\bibliography{ref}
\end{document}